\documentclass{aa}
\usepackage[varg]{txfonts}
\usepackage{graphicx}
\usepackage{xcolor}   
\usepackage{wasysym} 

\begin{document}
\title{WISE J005128.08+645651.7: An enigmatic R Coronae Borealis star}
\titlerunning{An enigmatic RCB star}
\author{J.~Budaj\inst{1} \and K. Bernhard\inst{2} \and I. Hubeny\inst{3} }
\institute{Astronomical Institute, Slovak Academy of Sciences,
05960 Tatransk\'{a} Lomnica, Slovak Republic\\
\email{budaj@ta3.sk}
\and 
Bundesdeutsche Arbeitsgemeinschaft f{\"u}r Ver{\"a}nderliche Sterne e.V. (BAV), Berlin, Germany 
\and
The University of Arizona, Steward Observatory, 933 North Cherry Avenue, Tucson, AZ 85719, USA}
\date{Received ???? ??, ????; accepted ???? ??, ????}

\abstract
{R Coronae Borealis (RCB) stars are hydrogen-deficient and carbon-rich (HdC) supergiants that likely originate from a merger of two white dwarfs. As such they are not expected to be members of close binary systems, in agreement with current observations.}
{We test the above hypothesis and search for eclipses of RCB stars that could be due to a companion.}
{We exploited old archival photometric data from Harvard photographic plates (DASCH) as well as modern-era photometric surveys such as Gaia, ASAS-SN, and ZTF to search for eclipses, and conducted follow-up spectroscopic observations.}
{We found one RCB star (WISE J005128.08+645651.7) that features two 3.5 yr long dimming events. We argue that these may be eclipses separated by an orbital period of 66 yr and that the star is an eclipsing binary candidate. The next eclipse is predicted for 2080. If confirmed, this would be the first RCB star in a binary system, the second-longest orbital period of any known eclipsing binary, and the longest duration of an eclipse. Light curve simulations support the idea that the eclipses could be caused by a dusty disk around the companion. 
The star shows unusually large pulsations. Our follow-up spectroscopy indicates a stellar temperature of about 11000\,K, the presence of sharp emission lines (including forbidden lines) during the normal brightness, and extremely strong diffuse interstellar bands. The star is likely surrounded by a hotter chromosphere and an extended low-density nebula. 
We calculated atmosphere models of the star. Models feature a density inversion and the derived chemical composition is compatible with some cool extreme He stars (EHe).
The star constitutes a bridge between the classical RCB, hot RCB, and EHe stars.}
{Future research is encouraged to confirm or refute the binary nature of the star and search for other RCB binaries.}

\maketitle
\nolinenumbers

\section{Introduction}

R Coronae Borealis (RCB) stars are a type of hydrogen-deficient carbon (HdC) star \citep{clayton96,clayton12}.
Such stars show no or only weak hydrogen lines and absorption features of \ion{C}{I}, CN, CO, or C$_{2}$.
However, helium is by far the most abundant element in their atmospheres \citep{asplund00} and many RCB stars
feature \ion{He}{I} 10830 emission or absorption features.
Their prototype, R CrB, is one of the first variable stars known and has been studied for more than two centuries \citep{pigott97}.
Nevertheless, these stars are very rare and their numbers increased significantly only very recently. 
As of 2025 we only know of about 162 RCB stars \citep{crawford25}. \cite{tisserand20} estimated that there should be no more than 500 RCB stars in the Galaxy.
RCB stars are supergiants with F-G spectral types, temperatures of about 5000-8000 K, and absolute magnitudes of $-5\leq M_{\rm V} \leq -3.5$ mag.
They are characterized by extreme irregular fadings (by up to 9 magnitudes in the V band), which is due to ejections and subsequent carbon dust condensation. These dimming events have a typical “reverse nova-like” shape and last from weeks to years before the star gradually returns to its original brightness as the dust dissipates.
In connection with that, all RCB stars feature an infrared (IR) excess caused by a warm dust with temperatures of about 400-1000 K \citep{tisserand12}.
They also show pulsations with periods of 30-100 days and amplitudes on the order of 0.1 mag \citep{lawson96,percy23}.
There is a separate, even smaller group of about five stars labeled as hot RCB stars. These also feature RCB-like fadings but their temperatures are of about 15-20 kK. They have emission lines and low surface gravity but seem to be fainter in the V band ($M_{\rm V} \approx -3$ mag), and their radii almost compare to those of main-sequence stars of similar temperatures. They are quite heterogeneous. Some of them may be the progeny of cool RCBs as they shrink and evolve toward hotter temperatures and some resemble Wolf-Rayet central stars of planetary nebulae  \citep{demarco02,tisserand20,tisserand22}.
For completeness, one has to note that there are two other similar groups of stars called dustless HdC (dLHdC) stars and extreme helium (EHe) stars \citep{crawford23}. The dLHdC stars are another subgroup of HdC stars that are very similar to RCB stars but that do not show dust obscuration events or IR excess.
EHe stars are hotter relatives of HdC stars that show strong He lines but no evidence of dust.

The origin of such rare HdC stars is not well understood.
They are likely the product of a merger of two white dwarfs (WDs): a CO and a He WD, which is referred to as a double degenerate scenario \citep{webbink84,han98,clayton12}. The merger is followed by a short-lived cool supergiant phase. Their masses are expected to be in the range
of 0.6-1.1 $M_\odot$ \citep{tisserand22}. Pulsations of these stars can be used to curb the stellar masses and they do not seem to contradict the merger scenario \citep{wong24}. 
An alternative hypothesis is that at least some RCB stars may originate from a final helium shell flash in the central star of a planetary nebula \citep{renzini90,iben96}. 
A consequence of the merger scenario is that RCBs are not expected to be found in binary systems. Indeed, to our knowledge, no RCB binary stars are known \citep{rao93,jeffery20}. This is quite unfortunate because binary stars can provide the most reliable masses of the stars, which is a fundamental and decisive parameter in stellar evolution. However, one must admit that detecting an RCB star in a binary is not easy. Given the size of these stars, the orbits will have long periods with low amplitude radial velocity curves. RCB stars do show variability in radial velocities of about 10-20 km\,$s^{-1}$ \citep{lawson97,feast19}. However, this variability is mainly associated with pulsations or large turbulent elements rather than with stellar companions.

All of this motivated us to search for possible companions of RCB stars via eclipses. As these stars are supergiants, the orbital periods will be long and the eclipsing objects will have to be huge too.
There are objects such as this and $\epsilon$ Aur is a famous example. Although this is not an RCB star itself, it is in many respects similar. It is an F supergiant and also features pulsations with the period of about 70 days similar to RCB stars. The star shows regular peculiar eclipses every 27 yr \citep{ludendorff03}. The eclipses are about 0.8 mag deep, last for about 2 yr, and are slightly asymmetric, with a flat bottom or even with a mid-eclipse brightening. They are caused by a huge dark dusty disk \citep{kloppenborg10} harboring an unseen B-type star \citep{hoard10}. The mid-eclipse brightening is not due to an inclined disk, as was believed for a long time, but due to flared co-planar disk geometry and forward scattering on dust \citep{budaj11epsaur}.  The radii of the supergiant and the disk are about 240 $R_\odot$ and 2.8 au, respectively \citep{kloppenborg15}.
There are other examples of such long-period eclipsing binaries or stars eclipsed by the disk or ring-like structures such as EE Cep \citep{mikolajewski99}, $\eta$ Gem \citep{torres22}, Gaia17bpp \citep{tzanizakis23}, AZ Cas \citep{galan12}, OGLE-LMC-ECL-11893 \citep{dong14}, ASAS J140748-3945.7  \citep{mamajek12,kenworthy15}, OGLE-BLG182.1.162852 \citep{rattenbury15}, ASASSN-21js \citep{pramono24}, and ASASSN-24fw \citep{johantgen24,nair24,zakamska25,shah26}, the record holder being
TYC 2505-672-1, which is an M giant eclipsed by a huge disk with a diameter of about 1.7-6.3 au surrounding a hot source. Its orbital period is 69 yr and only two total eclipses (lasting 3.45 yr) were observed so far \citep{rodriguez16,lipunov16}. The beginning of the next eclipse is scheduled in 2080.
We found ourselves looking, therefore, for analogs of $\epsilon$ Aur among RCB stars in archival photometric data and we had to go into the distant past. We inspected about 234 stars that were classified as RCB stars or candidates in the VSX catalog. We found two eclipse-like events of one RCB star, which is the subject of this paper.

\section{WISE J005128.08+645651.7}
\label{star}

Our star (WISE J005128.08+645651.7) was classified as an RCB candidate by \cite{otero14},
who found an RCB-like fading event with a 1.5 magnitude drop around MJD\,51500 in the Northern Sky Variability Survey (NSVS) database \citep{wozniak04b} and semi-regular pulsations with a predominant period of about 30 d.
The ASAS-SN survey detected a 1.2 magnitude drop around MJD\,57100, which was interpreted as an RCB fading event, and the star was also classified as RCB: \citep{jayasinghe18}. \cite{tisserand20} included the star in their list of RCB candidates based on its IR excess and brightness. 
The RCB classification was confirmed by \cite{karambelkar21} based on the IR spectra. Its IR spectrum shows blueshifted helium absorption, no hydrogen absorption lines, and several emission lines. Similar to other hot RCB stars, the emission lines are double-peaked, which were interpreted as an equatorial outflow. However, this star does not show the optical emission lines seen in other hot RCBs. Authors state that it is hot and pulsating with a tentative period of about 30 days.
\cite{crawford23} noted that this star “appears to border between the warmest HdCs and the coolest EHe stars.” Their low-resolution spectra covering red part of the optical region resemble the HdC0 spectral class.

The basic properties of the star were taken from the {\it Gaia} Data Release 3 \citep{gaia23} and are summarized in Table \ref{tab:star}. Other designations are NSVS J0051273+645649, GSC 04025-00779, and IRAS 00483+6440. According to {\it Gaia}, the star is at the distance of about 6500 pc, close to the Galactic plane, in the constellation of Cassiopeia. The distance from \cite{bailer-jones21} is slightly smaller (5300 pc). 
As we shall see below, these distances are not very reliable, and we shall revisit the distance in Sec. \ref{sed}. The stellar field centered on the star is shown in Fig. \ref{fig:img}. The star is clearly seen in DSS2-blue and was barely detected in Galex in the near-ultraviolet (NUV) region. It becomes the dominant source in IR WISE passbands. Its RUWE value is 1.5, indicating departures from a single star solution usually caused by an unresolved companion.
To estimate the dust extinction toward the star, we consulted 3D dust extinction maps from \cite{green19}. The extinction at this distance and in this direction is significant and amounts to A(g)=3.4 mag. However, we shall revisit this extinction in Sect. \ref{sed} too.

\section{Photometry}
\label{photometry}

We compiled the archival photometric observations of this star in several passbands from different sources. Digital Access to a Sky Century at Harvard (DASCH) \citep{grindlay12} has the longest baseline of 82 yr from 1899 to 1982. It has recently provided the community with its data release 7 (DR7). These are data from the photographic plates corresponding to the B filter.
ASAS-SN \citep{shappee14,christy23} contains about two years of data in filter V covering the eclipse-like event. 
Similarly Gaia DR2 \citep{gaia16,gaia23} contains two years of data that cover the event in three filters: G, BP, and RP.  
Data from the Zwicky Transient Facility (ZTF) \citep{bellm19} have the best precision, extend over 6.5 yr, and come in two filters, ZTF-r and ZTF-g.
The ATLAS project \citep{tonry18,heinze18} contains 10 yr of data in two filters: ATLAS-o and ATLAS-c.
The Wide-field Infrared Survey Explorer (WISE) and NEOWISE \citep{wright10,mainzer11} provide
about 14 yr of data in two filters, W1 and W2, centered at about  3.4 and 4.6 micron, respectively.
For bright sources, the W1 and W2 magnitudes are affected by partial saturation and increased scatter. Following the official NEOWISE Explanatory Supplement, we applied the tabulated saturation correction by adding it to the measured magnitudes in order to obtain corrected values. The correction is magnitude-dependent and leads to a slightly flatter long-term brightness evolution in both passbands.
Palomar Gattini-IR (PGIR) \citep{murakava24} survey provides 4 yr of observations in the J band in the IR region. Northern Sky Variability Survey (NSVS) \citep{wozniak04} monitored this object for about 10 months and detected an RCB-like event. These data cover the optical region without any filter using a red sensitive CCD.

\begin{figure}[ht]
\centerline{
\includegraphics[width=0.49\textwidth]{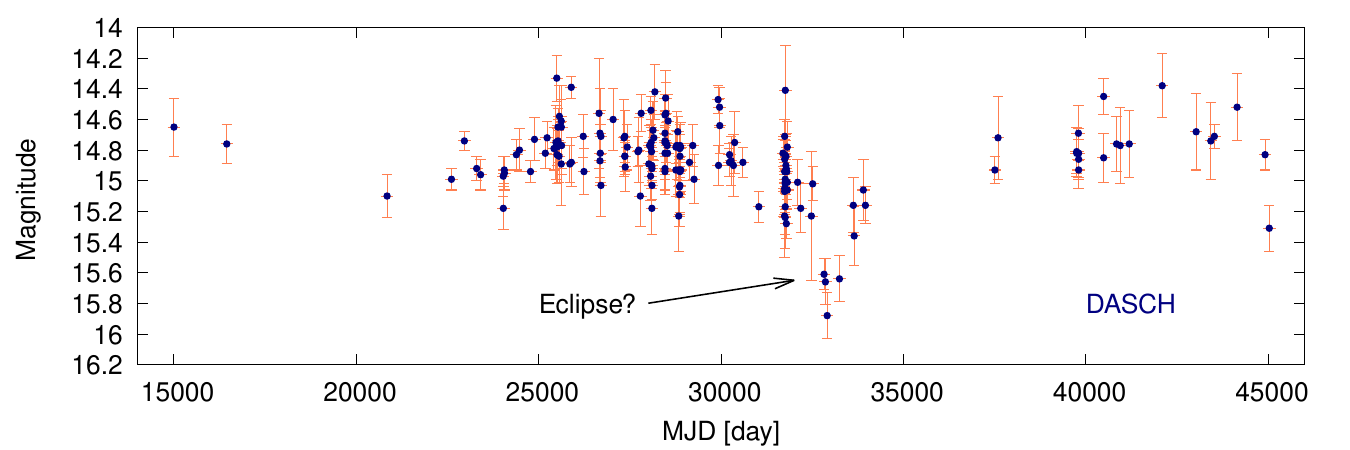}}
\caption{Historic light curve from DASCH photographic plates. An arrow points to the event that occurred in 1949 and that is most likely an eclipse.}
\label{fig:dasch}
\end{figure}

The DASCH data are displayed in Fig. \ref{fig:dasch}. The magnitudes and corresponding uncertainties were taken from the DASCH homepage. One can see a 1 mag deep drop in the brightness that occurred around MJD\,33000. All digitized plates in the vicinity of the event were visually inspected, as well as analyzed using the Aperture Photometry Tool (APT) version 3.0.9 \citep{laher12}. This resulted in two additional data points that were not detected by the automated object recognition but that could be evaluated using APT. 
Similarly, we added two further data points evaluated using APT near MJD\,45000. In Sect. \ref{eclipse} we argue that this drop in the brightness around MJD\,33000 is in fact an eclipse. The two latter data points near MJD\,45000 were used for the period analysis to exclude solutions corresponding to half the true orbital period.

\begin{figure}[ht]
\centerline{
\includegraphics[width=0.45\textwidth]{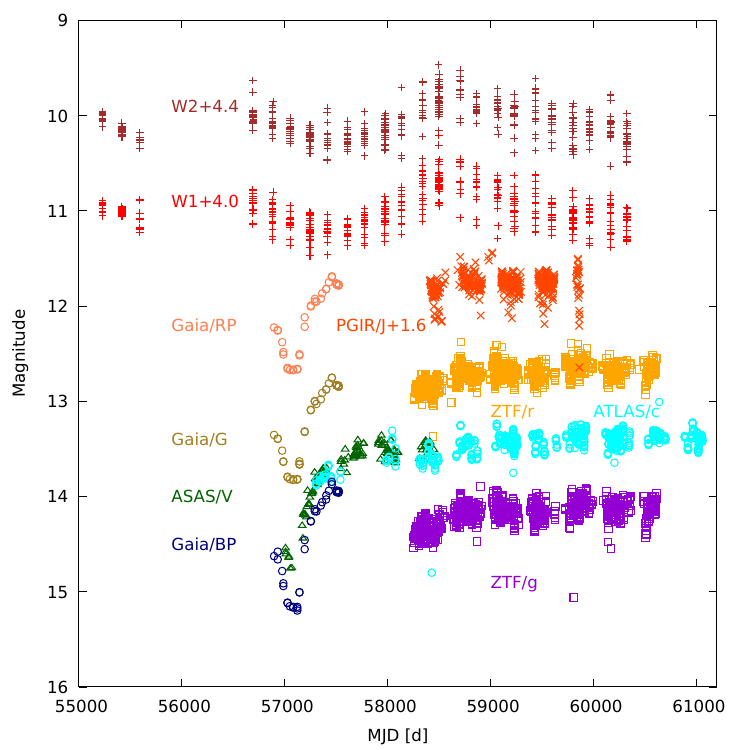}}
\caption{Modern era photometry of the star in different filters.
Some filters were offset for clarity and the shift in magnitudes is indicated in the label.}
\label{fig:all}
\end{figure}

Modern-era photometric data are shown in Fig. \ref{fig:all}. Observations in ATLAS-o filter are not shown there for clarity but are seen, for example, in Fig. \ref{fig:atlas}.
One can see the eclipse-like event at MJD\,57000 in the optical region that is described in more detail in Sect. \ref{eclipse}. 
Infrared data do not show the eclipse but a wave with a period of about 7 yr and an amplitude of about 0.5 mag.

\begin{figure}[ht]
\centerline{
\includegraphics[width=0.45\textwidth]{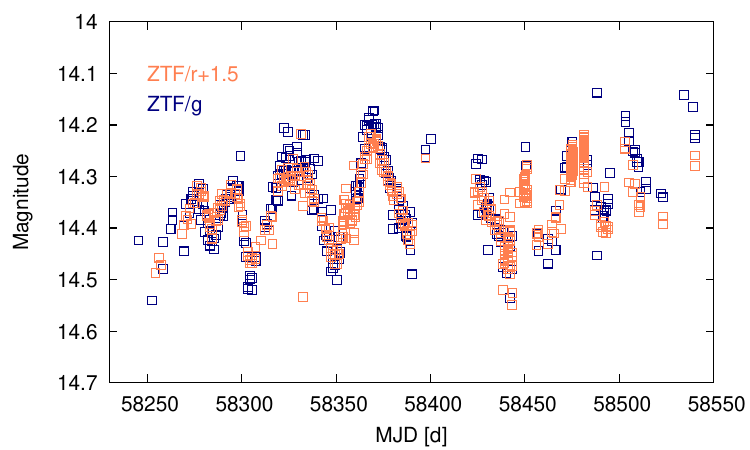}}
\caption{Pulsations seen in ZTF-r and ZTF-g data. Magnitudes in ZTF-r were offset by 1.5 mag to enable an overlay with the ZTF-g data.}
\label{fig:pulsation}
\end{figure}

Apart from the eclipse-like event one can observe faster semi-regular variability with the amplitudes up to 0.3 mag and periods from 20 to 40 days mentioned already by \cite{otero14}. This is best seen in ZTF data and shown in Fig. \ref{fig:pulsation}. The amplitude in ZTF-g seems slightly higher than in ZTF-r.
This kind of variability is most likely due to pulsations and is typical of RCB stars.
An analysis of the ZTF-g data using the Lomb–Scargle GLS method in \textsc{Peranso} \citep{paunzen16} reveals semi-regular pulsations with a dominant period of about 33.6 days and a peak-to-peak amplitude sometimes reaching up to 0.4 mag. Such an amplitude is unusually large, but it does not  exceed those of RY~Sgr, which shows the largest known pulsational amplitude among hydrogen-deficient stars.
The dominant period of pulsations of RY~Sgr is 37.79 days \citep{crause07}, with a variable semi-amplitude of the corresponding best-fit sinusoid of about 0.05–0.25 mag according to \cite{percy18}. The ASAS-3 $V$ data \citep{Pojmanski02} and AAVSO observations \citep{AAVSO_DB} also reveal individual pulsation cycles with larger peak-to-peak amplitudes, reaching up to 1 mag in $V$, which reflects both the doubling from semi- to full amplitude and the presence of additional variations.

\section{Eclipses}
\label{eclipse}

\begin{figure}[ht]
\centerline{
\includegraphics[width=0.45\textwidth]{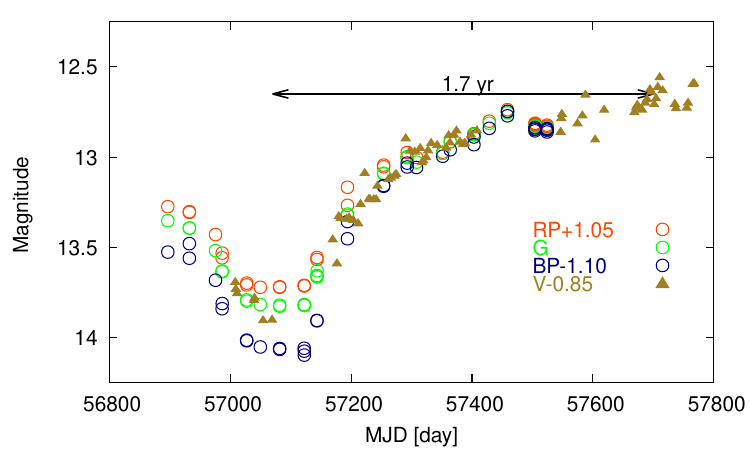}}
\caption{Gaia eclipse in three Gaia and one ASAS-SN passbands.
Values in different filters were offset to match the egress wing of the ASAS-SN-V filter and the offset is indicated in the label.}
\label{fig:eclipse}
\end{figure}

The latest eclipse-like event was recorded in Gaia and ASAS-SN data. It is shown in Fig. \ref{fig:eclipse}. We refer to it also as the “Gaia eclipse” to distinguish it from the “DASCH eclipse” detected in DASCH data. As can be seen, these data cover most of the eclipse but
part of the ingress is missing. Assuming that the eclipse is symmetric, the whole eclipse would last about 3.5 yr, which is comparable to the eclipse duration of TYC 2505-672-1 \citep{rodriguez16}. 
The eclipse features a flat-bottomed part that lasts for about 3.5 months. It is also similar to  TYC 2505-672-1 and it indicates a total eclipse during which one of the components is fully eclipsed. Alternatively, such a flat-bottomed eclipse might be caused by a nonspherical body such as an edge-on disk, which does not have to  occult the whole star, as in the case of $\epsilon$ Aur \citep{budaj11epsaur,kloppenborg15}.
The depth of the eclipse varies with wavelength and is deeper at shorter wavelengths
(1.4 mag in the Gaia BP filter versus 1.1 mag in the Gaia RP filter). This indicates that it is the hotter and smaller but more luminous component that is eclipsed during this eclipse.  If the eclipse were total, it would result in the luminosity ratios of the hotter and cooler objects of about 1.7 and 2.6 at RP and BP passbands, respectively. However, as we shall see below in Sect. \ref{RCB_fading}, the eclipse is most likely not a total eclipse.

\begin{figure}[ht]
\centerline{
\includegraphics[width=0.45\textwidth]{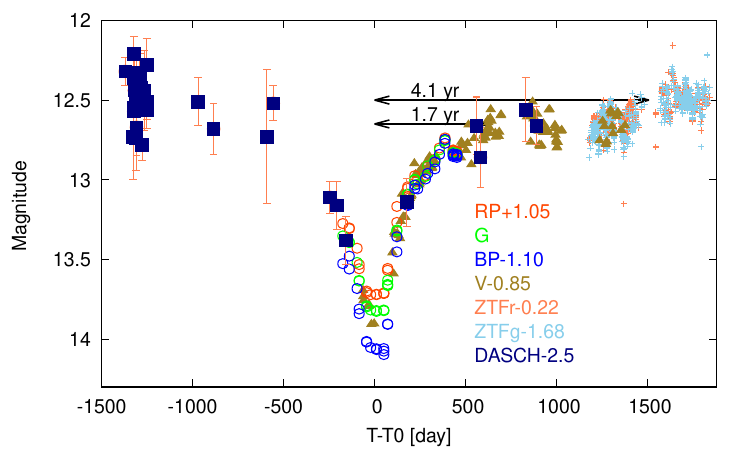}}
\caption{Gaia and DASCH eclipses phased with the orbital period of 65.7 yr. 
For clarity, only the DASCH data contain error bars.
Values in different filters were offset to match the egress wing of the ASAS-SN-V filter.}
\label{fig:eclipse_phase}
\end{figure}

The center of the Gaia eclipse is at MJD\,57071(2), which is February 2015.
If we assume that the DASCH event on MJD\,33000 is also an eclipse and phase both eclipses together then we get an orbital period of 24006(2) days, i.e., 65.7 yr, and the center of the DASCH eclipse at MJD\,33065, which is May 1949. Consequently, the center of the next eclipse is predicted on MJD\,81077, which is November 2080, but the beginning of eclipse would be about 2 yr earlier. Coincidentally, this is also in the same year as  TYC 2505-672-1.  The orbital period of 65.7 yr is very long. In fact, it would be the eclipsing binary with the second longest orbital period known, after TYC 2505-672-1 (69 yr), but before 
$\epsilon$ Aur (27 yr). 
In Fig. \ref{fig:eclipse_phase} we show the two eclipses phased with the above-mentioned period.
DASCH data seem to match the Gaia eclipse very well including its duration and out-of-eclipse brightness. That is why we think the DASCH event is another eclipse.
The ephemeris for the minima is
\begin{equation}
 MJD=57071(2) + 24006(2)\times E.
\end{equation}
However, note the behavior of ZTF data in Fig. \ref{fig:eclipse_phase}. They indicate that the egress part of the eclipse may be even longer, about 4 yr, which would make the disk significantly bigger.

To investigate whether fractional periods — specifically p/2, p/3, p/4, and p/5 — could be plausible, the digitized DASCH plates from the relevant epochs were visually inspected and analyzed with APT, similar to the eclipse event around MJD\,33000. As a result, periods from p/2 to p/5 can be excluded. Even shorter periods (<11 years) are ruled out by modern photometric data of Gaia, ASAS-SN, and ZTF.
The two (and only) data points near a possible secondary eclipse at phases 0.493 and 0.498
(MJD\,44912, 14.83 mag and MJD\,45026, 15.31 mag, respectively) may indicate a shallow secondary eclipse with an amplitude of about 0.2–0.3 mag.
However, the evidence of a secondary minimum is currently insufficient to draw a definitive conclusion.
Apart from the two above-mentioned primary eclipses and a possible shallow secondary minimum, we do not see any evidence of further dimming events in the available data, with the exception of the event at MJD\,51500 described in Sect. \ref{RCB_fading}.

\section{RCB fading event}
\label{RCB_fading}

\begin{figure}[ht]
\centerline{
\includegraphics[width=0.45\textwidth]{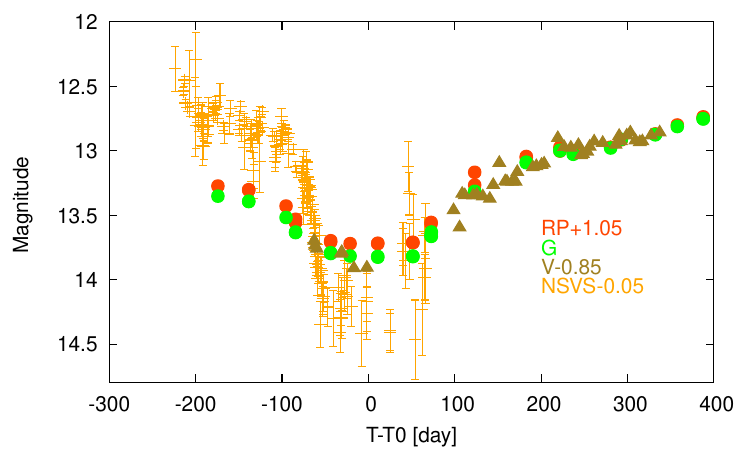}}
\caption{RCB fading in NSVS data that occurred in 1999 (in orange with errors) in comparison with the Gaia eclipse in 2015 (Gaia/RP-red, Gaia/G-green, and ASAS-SN/V-olive). Values in different filters were offset by the magnitude values indicated in their labels.}
\label{fig:eclipse_comparison}
\end{figure}

Apart from the two potential eclipses, our star shows fading events that are typical distinguishing characteristics of the RCB phenomenon. Unlike eclipses, RCB fadings usually feature much more complicated shapes with a steep ingress followed by a more gradual egress, often with additional multiple fadings \citep{schaefer04,crawford25}. 
One such event, recorded in NSVS and already noticed by \cite{otero14}, occurred in 1999 at MJD\,51500 and is shown in Fig. \ref{fig:eclipse_comparison} in comparison with the Gaia eclipse.
The data show a brightness drop of about 1.6 mag with a flat bottom, making it the deepest feature recorded for this star.
Its ingress is much steeper than that of the eclipse and its depth is 0.5 mag deeper than the eclipse in the Gaia RP filter.
NSVS uses broadband CCDs that are sensitive in the red optical region similar to the Gaia RP passband.
This deep fading event means that the eclipse event is not a total eclipse, i.e., the RCB star is not fully eclipsed during the eclipse.
Assuming that we have a binary star with the 1.6 mag fading and that the primary RCB component is fully hidden during the fading, one could estimate the maximal contribution of the secondary to the total light, $I_2/(I_1+I_2)\approx 0.23$, or a minimal luminosity ratio of the two stars in the NSVS passband of about $I_1/I_2=3.4$ .

\section{SED}
\label{sed}

A spectral energy distribution (SED) can shed light on the nature and properties of the star.
We collected photometric observations of the star in different filters using the VOSA tool \citep{bayo08}. In addition, the star was marginally detected by GALEX \citep{morrissey07} in the NUV band. Using the gPhoton database of GALEX photons \citep{million16}, we estimated its brightness in this band as a 21.0 ± 0.3 mag source.
Furthermore, we downloaded the SWIFT images \citep{roming09} and analyzed them using the Aperture Photometry Tool \citep{laher12}. As a result, we estimated the brightness of the star in the UVW2 passband to be $22.1\pm0.4$ mag.
Unfortunately, the SWIFT image was taken just during the Gaia eclipse. This means that the UVW2 brightness out of the eclipse is likely at least 1.4 mag (which is the depth of eclipse in Gaia BP passband) brighter. We then added these two points manually to the SED. 

\begin{figure}[h]
\centerline{
\includegraphics[width=0.45\textwidth]{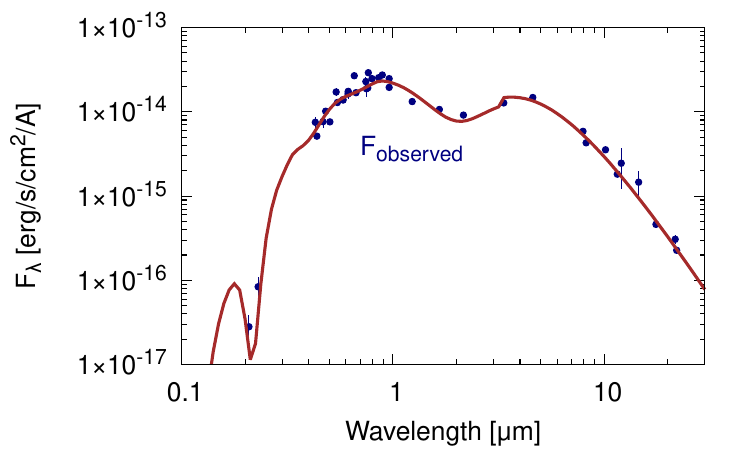}}
\caption{SED of the star. Photometric observations are in blue. 
A two black body fit with the reddening is in brown. }
\label{fig:sed2}
\end{figure}

The observed SED is displayed in Fig. \ref{fig:sed2}.
It clearly shows two peaks: one in the optical and the other in the IR region referred to as an IR excess, which is typical for RCB stars.
The star is experiencing brightness fluctuations and as a result of that the observed error bars do not represent the real uncertainties very well.
The Gaia distance to the star is not very reliable.
\cite{tisserand24} found that in RCB stars that had declines during the Gaia observations, such as our star, the astrometric fits are not valid due to the Gaia point spread function chromaticity effect in both shape and centroid.
For these reasons we refrained from using the Gaia distance and the extinction from \cite{green19}.

To get a more reasonable estimate of these parameters, we tried a slightly different approach. Based on absolute magnitudes of RCB and hot RCB stars, we assumed an absolute magnitude of the RCB star of $M_{\rm V}=-4\pm 0.5$ mag. Using a bolometric correction of $BC=-0.44$ mag \citep{pecaut13}, one gets a bolometric absolute magnitude of $M=M_{\rm V}+BC=-4.44$ mag. Adopting a temperature of the star obtained in Sec. \ref{mod}
from the model atmospheres and spectroscopy, which is about $T_{\rm eff}=11\,000$\,K, one gets a star radius of $R=19\pm 5\, R_{\odot}$. We then created a two black body model with the extinction according to \cite{cardelli89} and fit it to the observed photometry. The distance, extinction, and parameters of the second black body were the free parameters, while the radius and temperature of the star (first black body) were fixed.
Due to the variability of the source, we added a uniform relative error (11-17\%) to each point in quadrature to the observed errors to ensure that the reduced $\chi^2$ is about one.
The SED fitting resulted in a distance of 5100$\pm$1200 pc, an extinction of E(B-V)=1.29 mag ($A_{\rm V}=4.00$ mag assuming $R_{\rm V}=3.1$), and the following parameters of the second black body: 
$T_2=750\,{\rm K}, R_2=1500\,R_{\odot}, L_2=700\,L_{\odot}$.
The fixed parameters of the first black body were
$T_1=11\,000\,{\rm K}, R_1=19\,R_{\odot}, L_1=4700\,L_{\odot}$.
The fit is shown in Fig. \ref{fig:sed2}. 
The extinction in the V band is one magnitude stronger than derived from the 3D maps of \cite{green19}, which predict essentially constant extinction beyond 6500 pc.
The first black body represents the RCB star. The second black body represents the dust envelope around the RCB star. It is huge (7 au) and its temperature is in agreement with IR excess in RCB stars that is caused by warm dust.\footnote{This second black body should not be confused with the dusty disk that we invoke in Sec. \ref{model} to produce eclipses. Such disk will be too small and too cold to be seen in the SED.}
Note that such black body fitting presumes an optically thick “solid surface” sphere. In reality, the dust is likely to be optically thin, at least at the edges, which means that the real extent of the dust envelope must be even bigger. 
If it were not spherical, but for example a disk-like structure, then its radius would be even bigger, a few tens of astronomical units. 
A saddle point between the black bodies is at about 2 micron, indicating that an
eclipse of the small optical source at the shorter wavelengths will not affect the flux from the extended IR source at wavelengths beyond 2 micron. The existence of this saddle point and the fact that the IR excess is well reproduced with a single black body indicates that this dust cloud is fairly uniform and lacks smaller and hotter particles as well as large cool grains (at large distances).

The IR excess can be used to estimate the minimum amount of dust necessary to produce it.
The amount of dust depends on the dust cloud geometry, its temperature, chemical composition, particle size, and optical depth; therefore, it is a highly degenerate problem. If the dust were optically thick, one could estimate its temperature and surface area, but not its mass (similarly to the black body fitting applied above). However, assuming that the dust is optically thin would make IR excess independent of the geometry and dependent on the dust mass, which would provide a lower limit on the amount of dust. 
We used a simple radiative transfer code Shellspec \citep{budaj19} to fit the IR excess. We assumed a distance of 5100 pc, E(B-V)=1.29 mag, $T_1=11000\,{\rm K}, R_1=19\,R_{\odot}$, and that the dust is made of spherical homogeneous carbon grains of various sizes ranging from 1 to 30 micron. The carbon dust opacities were taken from \cite{budaj15}. We did not consider smaller particles because they are optically active mainly in the optical region and would strongly attenuate the light from the star. 
The lower limit on the total amount of carbon dust we obtained is $1\,10^{-8} M_{\odot} (3\,10^{-3} M_{\oplus})$ for 1 mic particles and $5\,10^{-7} M_{\odot} (0.2 M_{\oplus}$) for 30 mic particles.

\section{Spectroscopy}
\label{spectra}

We observed the star with 2.56 m NOT telescope located at Canary Islands over three nights. More detailed information about the spectra can be found in Table \ref{tab:spectra}.
We used FIES \citep{telting14}, a high-resolution Echelle spectrograph in its lower-resolution mode with a fiber diameter of 2.5" and with 1x1 binning.
Each night we obtained 3x15 min exposures of the target plus standard ThAr and flat-field frames. During the observations the airmass was about 1.3-1.5.
The spectra were reduced using the automatic pipeline. Spectra from the first two nights do not differ significantly so all six exposures were subsequently co-added into one spectrum, removing the cosmic rays, and corrected for an averaged heliocentric velocity of about +13.6 km\,s$^{-1}$. This first spectrum has maximum S/N $\approx 40$, R=25000, and covers the interval 3820-9120 \AA. The three exposures from the last night were co-added separately into a second spectrum in a similar way and corrected for an averaged heliocentric velocity of +3.2 km\,s$^{-1}$ .

\begin{figure}[ht]
\centerline{
\includegraphics[width=0.45\textwidth]{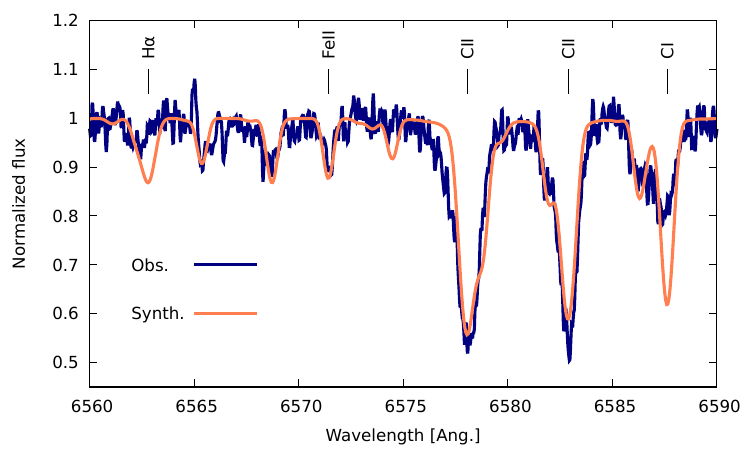}}
\caption{Strong \ion{C}{II} lines and an absence of hydrogen lines confirm that this is a hot HdC star. The spectrum was shifted to the laboratory frame and the synthetic spectrum is shown for comparison.}
\label{fig:cii}
\end{figure}
\begin{figure}[ht]
\centerline{
\includegraphics[width=0.45\textwidth]{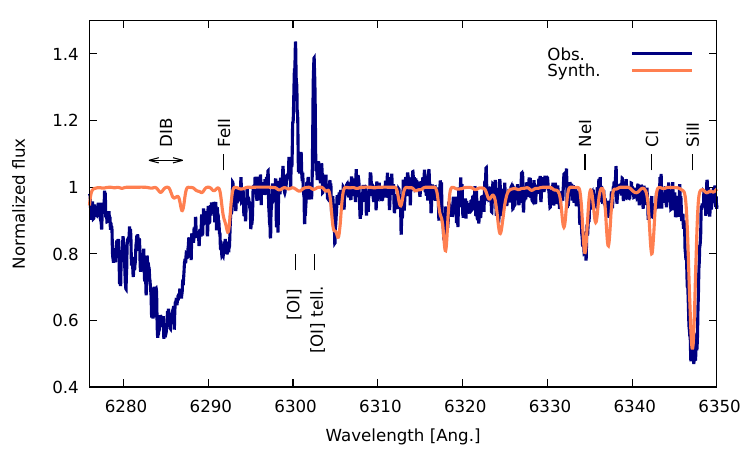}}
\caption{Spectrum features strong and broad DIBs as well as two forbidden [OI] 6300 emission lines (the one shifted to the blue is from the star, the sharper one shifted to the red is telluric). The spectrum of the star was shifted to the laboratory frame and the synthetic spectrum is shown for comparison.}
\label{fig:dib}
\end{figure}

The first spectrum of the star shows strong absorption lines of \ion{He}{I}, \ion{C}{I-II}, \ion{N}{I-II}, \ion{O}{I}, \ion{Ne}{I}, \ion{Mg}{II}, \ion{Al}{II}, \ion{Si}{II}, \ion{S}{II}, \ion{Fe}{I-II}, and \ion{Ca}{II}. Some of these lines were already identified in the low resolution optical spectrum of the star obtained by \cite{crawford23}. A few selected lines are listed in Tables \ref{tab:lines1} and \ref{tab:lines2} together with their equivalent widths and radial velocities. The radial velocity of the star from these lines is about $-90.8\pm 0.3$ km\,s$^{-1}$, which is quite high for the star close to Galactic plane.
Hydrogen lines are not detected (see Fig. \ref{fig:cii}), which confirms that this is a hydrogen-deficient star. The presence of a fading event, IR excess, and high luminosity ranks this star among RCB stars. The presence of \ion{He}{I}, \ion{Ne}{I}, and strong \ion{C}{II} lines confirms the findings of \cite{crawford23} that this star is a transition between the hottest HdC and coolest EHe stars. At the same time, it is significantly cooler than a small group of hot RCB stars \citep{demarco02,tisserand20}. 

Apart from the stellar lines we observe forbidden telluric emission lines ([OI] 5577, 6300, and 6364) and  interstellar \ion{Na}{I}\,D absorption lines. Surprisingly, we also observe forbidden [OI] 6300 emission from the star, which indicates the presence of a weakly ionized low density oxygen-rich nebula around the star (see Fig. \ref{fig:dib}). The emission line is narrow and dominated by the instrumental broadening. It means that the rotation or expansion velocity of the nebula cannot be larger than 18 km\,s$^{-1}$.
A similar but broad forbidden emission lines are sometimes seen in RCB stars during deep declines. Hot RCB stars, with temperatures of about 15-20 kK, do show forbidden emission lines during maximum light \citep{demarco02}.  Their forbidden lines are mainly lines from ionized species originating in the ionized low-density nebula. Hence our star, which is cooler and which features neutral forbidden lines, is bridging the properties of RCB and hot RCB stars. Namely, the nebula properties may represent an “evolution” sequence ranging from almost invisible in RCB stars (visible only when the star is attenuated) to becoming visible during maximum light in neutral forbidden lines of our star, toward more ionized nebulae of hot RCB stars.

Finally, one can see several wide and strong diffuse interstellar bands (DIBs, see Fig. \ref{fig:dib}). Actually, these are some of the strongest DIBs ever observed \citep{cox05,luna08}. They may be caused by large carbon-bearing molecules \citep{campbell15} but their exact origin is not known, so this star may help to shed the light on this mystery too.
The equivalent widths of these DIBs are listed in Table \ref{tab:dib}. The strengths of the DIBs correlates with the presence of dust in the interstellar medium. That is why they can be used to estimate the interstellar extinction.
We used the calibration of \cite{luna08} and the resulting E(B-V) values are also listed in Table \ref{tab:dib}.
All three weaker DIBs give consistent values of about 0.90-0.96, which are in excellent agreement 
with the values we found previously based on the distance and 3D extinction maps of \cite{green19}.
However, the two strongest DIBs (5780 and 6284) give significantly stronger extinction, which would result in much higher temperature and luminosity of the RCB star. A possible explanation is that DIB 5780 is too strong and is in the extrapolation regime of \cite{luna08} calibration and the DIB 6284 is affected by telluric lines. Alternatively, these DIBs might also be affected by the circumstellar dust with different properties than the interstellar dust because our star is an RCB star. 
Sodium \ion{Na}{I}\,D lines might also be used to trace the interstellar absorption.
These lines are very strong and have very complicated shapes with about five blueshifted absorption components
located at about -13, -46, -69, -92, and -126 km\,s$^{-1}$. Their EQWs are also listed in the Table \ref{tab:dib}.
Using the calibration of \cite{poznanski12}, one would get unrealistically high values of extinction from these lines.
There are also three sharp absorption lines at about 7662-7765\AA\ and three similar absorption lines increasing in strength at about 7696-7699\AA. These are most likely three components of the \ion{K}{I} 7665, 7669\AA\ resonance doublet, which have similar properties to the three strongest components of \ion{Na}{I}\,D lines.
\begin{table}[ht]
\caption{
DIBs and \ion{Na}{I}\,D lines.}
\label{tab:dib}
\centering
\begin{tabular}{lrr}
\hline\hline
Wavelength  & EQW  & E(B-V) \\
\hline
5779.70     & 1.20 &  2.61 \\
5796.83     & 0.16 &  0.94 \\
6283.16     & 1.79 &  1.99 \\
6613.35     & 0.19 &  0.90 \\
7223.64     & 0.24 &  0.96 \\
NaI\,D$_2$5890 & 1.86 & - \\
NaI\,D$_1$5896 & 1.68 & - \\
\hline
\end{tabular}
\tablefoot{The columns are the wavelength, the equivalent widths (both in Angstr\"oms), and E(B-V) in magnitudes  based on the calibration of \cite{luna08}.}
\end{table}
\begin{figure}[ht]
\centerline{
\includegraphics[width=0.45\textwidth]{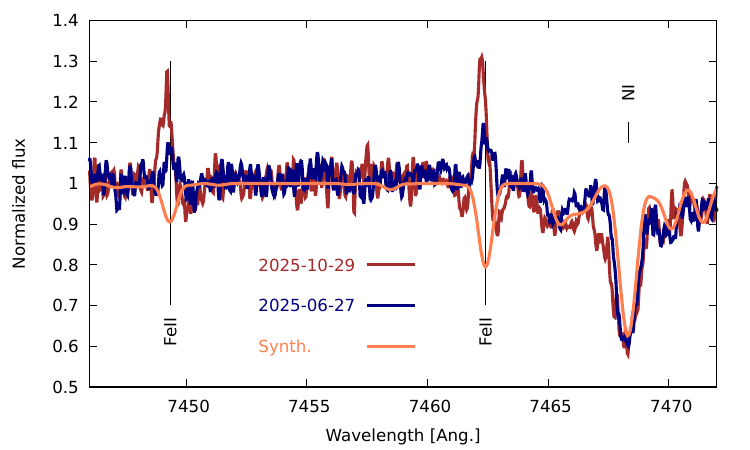}}
\caption{Second spectrum (2025-10-29) features sharp \ion{Fe}{II} emission lines that are slightly shifted to the blue and superimposed on the underlying broader absorption lines.
The spectra of the star were shifted to the laboratory frame and the synthetic spectrum is shown for comparison.}
\label{fig:feii}
\end{figure}
\begin{figure}[ht]
\centerline{
\includegraphics[width=0.45\textwidth]{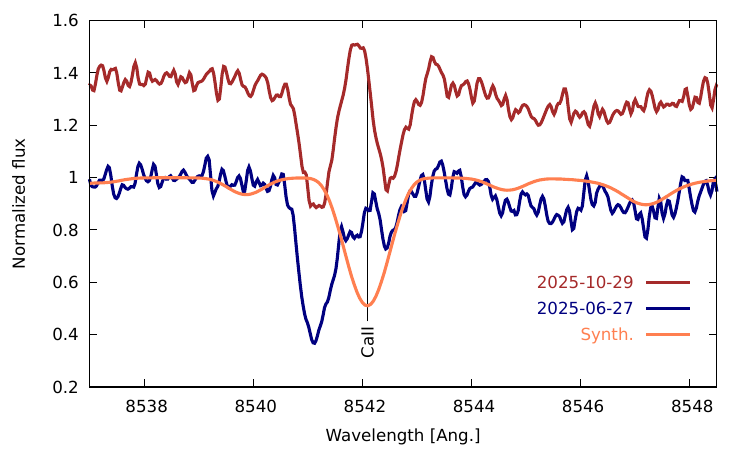}}
\caption{Second spectrum (vertically offset for clarity) also features sharp \ion{Ca}{II} emission lines that are slightly shifted to the blue and superimposed on the underlying broader absorption with a strong blue wing. The spectra of the star were shifted to the laboratory frame and the synthetic spectrum is shown for comparison.}
\label{fig:caii}
\end{figure}
\begin{figure}[ht]
\centerline{
\includegraphics[width=0.45\textwidth]{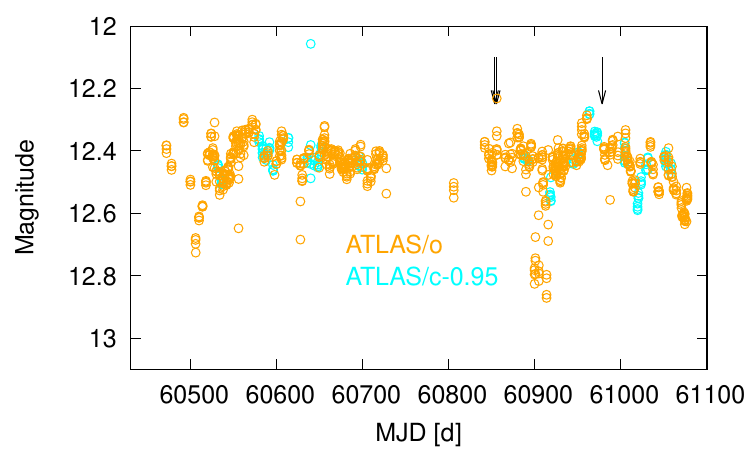}}
\caption{Brightness of the star in ATLAS-c and ATLAS-o filters during the time when the spectra were taken (black arrows). Values in ATLAS-c filters were offset and the shift in magnitudes is indicated in the label.}
\label{fig:atlas}
\end{figure}

The second spectrum obtained four months later is similar to the first spectrum described above. However, now we identified numerous sharp emission lines; for example, at 5325, 5991, 6238, 6247, 6416, 6432, 6516, 7449, 7462, 7515, 8402, 8433, and 8468 \AA. These are mainly low-excitation \ion{Fe}{II} lines (see Fig. \ref{fig:feii}). Their radial velocities and equivalent widths are listed in Table \ref{tab:emlines}. It seems that their strength increases with the wavelength and their average radial velocity is $-98.1\pm 0.5$ km\,s$^{-1}$. All these lines coincide with the radial velocity of the star but are slightly shifted to the blue. Their half-width is not larger than 18 km\,s$^{-1}$ and is mainly determined by the spectral resolution or instrumental broadening. They likely originate from the denser partially ionized regions close to the star or a sort of a chromosphere (temperature inversion). The increased emission is sometimes accompanied by an occurrence of an underlying broader absorption. This indicates some phenomenon cooling the photosphere and transferring the energy to the slowly expanding chromosphere. Again, such emission lines are usually seen in RCB stars during deep declines but here we observe them during the maximum light. Hot RCB stars may show emission lines at maximum light, but they are usually stronger, broader, double-peaked \ion{C}{II}, \ion{He}{I}, \ion{O}{I}, and \ion{Fe}{II} lines that likely originate from a disk wind \citep{demarco02,tisserand20}. Hence, our star with transient, weak, and sharp \ion{Fe}{II} emission may represent an evolutionary link between RCB and hot RCB stars. The brightness of the star is illustrated in Fig. \ref{fig:atlas}. There is no significant fading of the star or significant pulsation activity at the moment when the second spectrum was taken. However, there might have been a weak dimming event between the spectra seen in ATLAS-o filter. In the future, it might be interesting to study these lines and their variability in more detail to see whether they are associated with some other phenomena. Similar sharp emission features were detected in the cores of broader \ion{Ca}{II} IR absorption lines, which also show an enhanced absorption in the blue wing as seen in Fig. \ref{fig:caii}. 
In addition to the above-mentioned [OI]\,6300, in this second spectrum we also detected a weak forbidden emission line [OI]\,5577 from the star.
Apart from that, some blended absorption spectral features at 5018, 5169, 6148, 6157, and 6455 \AA\ significantly changed their shapes.

\section{Model of the atmosphere}
\label{mod}

\begin{figure}[ht]
\centerline{
\includegraphics[width=0.45\textwidth]{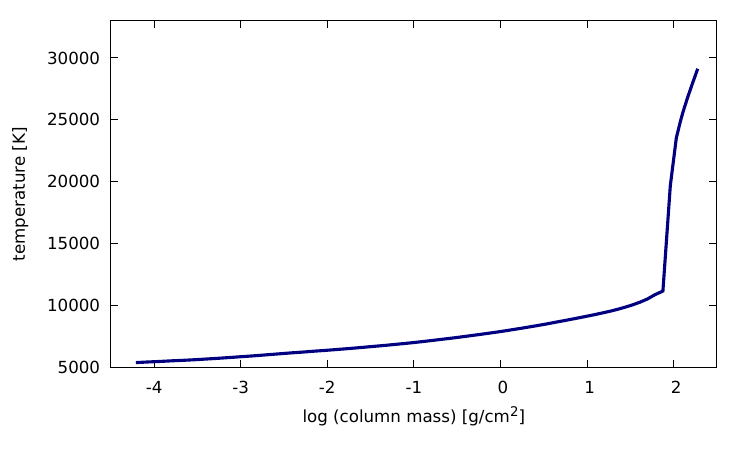}}
\centerline{
\includegraphics[width=0.45\textwidth]{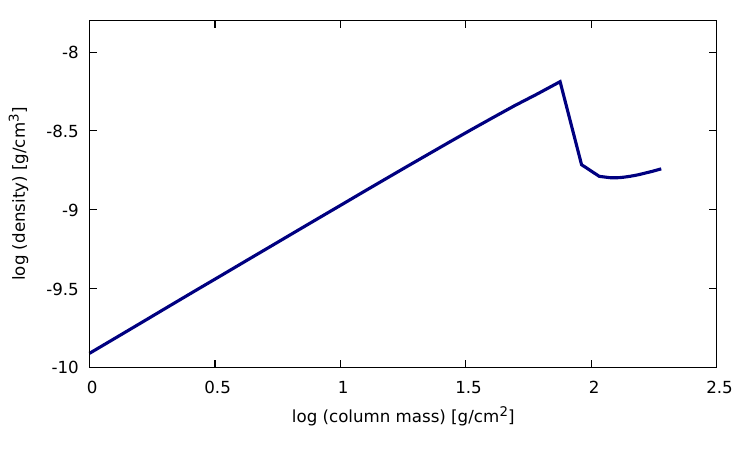}}
\caption{
Atmosphere models. The temperature structure is in the upper panel. A close look at the density inversion is in the lower panel.}
\label{fig:t_mass}
\end{figure}

To estimate the properties of the star, such as its temperature and chemical composition, we calculated a sequence of atmosphere models using the code TLUSTY \citep{hubeny88,hubeny21}. 
This is a state-of-the-art code well known for not being restricted to the local thermodynamical equilibrium (LTE). However, it has never been used to calculate atmosphere models of such HdC supergiants. That is why we briefly describe the basic assumptions and setup. We used the version 209v, which was modified by IH to model hydrogen-deficient, helium- and metal-rich atmospheres of WDs \citep{budaj22}. These are 1D models assuming hydrostatic, radiative, and convective equilibrium. It was difficult to converge the models, which is why we restricted ourselves to LTE only. We assumed line blanketing from iron lines and continuum opacities arising from these ions: 
\ion{H}{I}, \ion{He}{I}, \ion{C}{I-II}, \ion{O}{I-II}, \ion{Mg}{II}, \ion{Al}{II-III}, \ion{Si}{II-III}, and \ion{Fe}{I-III}. The convective transport of energy was not very strong, which is why we turned it off completely. Molecules were not considered either. The optically thick part of the atmosphere features a density inversion (see Fig.\ref{fig:t_mass}).
It was noticed before by \cite{asplund97} and was associated with the ionization of He and the resulting decrease in the mean molecular weight. This density inversion may be driving pulsations via the so-called strange mode instability \citep{gautschy90}.
Our 1D static atmosphere models only represent an approximation of the true behavior of the star, which will require a more detailed analysis in the future, but we feel that they are sufficient for the purpose of the present study.
 
\begin{figure}[ht]
\centerline{
\includegraphics[width=0.45\textwidth]{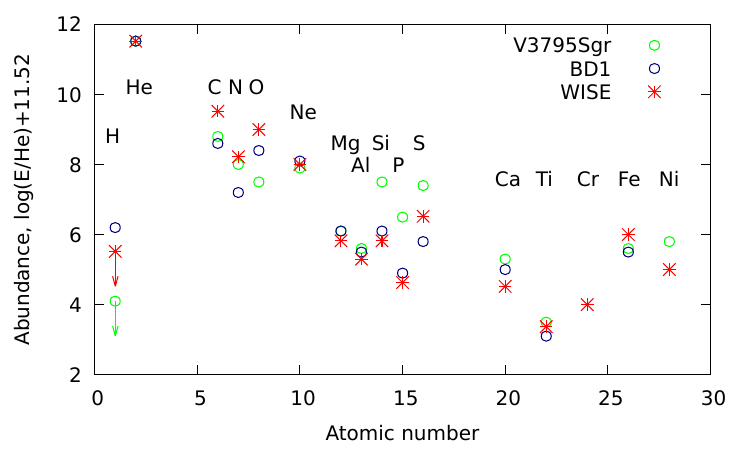}}
\caption{
Abundances of our star (red asterisks) compared to the hottest RCB and the coolest EHe stars (circles) from the sample of \cite{asplund00}.}
\label{fig:abn}
\end{figure}

Synthetic spectra from the atmosphere models were calculated using the code SYNSPEC \citep{hubeny21}. These were compared with the observed spectra and new atmosphere models and new spectra were calculated and iterated to fit the observations changing the abundance of one chemical element at the time.  In this way both the synthetic spectra and the atmosphere models are consistent with the chemical composition. Taking into account the instrumental broadening, we estimated the projected rotational velocity, $v\sin i$, of the star of about 10-20 km\,s$^{-1}$.
Based on the synthetic spectra, the effective temperature of the star is approximately 10000-12000\,K 
(see e.g., a comparison of our \ion{C}{I-II} lines from Fig. \ref{fig:cii} with Fig. 1 of \cite{pandey01}).
Our final model has $T_{\rm eff}=11200$ K, $\log g=1.3$ (cgs) K, and microturbulence of 8 km\,s$^{-1}$ \citep{asplund00}. The chemical composition corresponding to this model is given in Table \ref{tab:chem} in the form of an abundance of some element, $E$, defined as number densities relative to helium:
\begin{equation}
    A(E)=\log \frac{N(E)}{N(He)}+12 .
\end{equation}
Note that abundances of RCB stars, $A(RCB)$, are often given in a slightly different scale. Assuming that the mass fraction of hydrogen is negligible, these abundances can be expressed as
\begin{equation}
    A(RCB)=\log \frac{N(E)}{N(He)}+11.52 .
\end{equation}
The scale is shifted by about 0.48 dex. This unconventional scaling is motivated by the assumption that RCB stars originated from initially normal stars with approximately solar chemical composition. Nuclear reactions changed their chemical composition, but the total number of nucleons remained conserved. We can compare the chemical composition of our star with the chemical composition of the hottest RCB star V3795\,Sgr ($T_{\rm eff}=8000$\,K) and the coolest EHe star BD1$^\circ$4381 ($T_{\rm eff}=8500$\,K) from the sample of \cite{asplund00}. The abundances are displayed in Fig.\ref{fig:abn}. As can be seen, our star is slightly more enriched in CNO elements, but its overall chemical composition resembles that of BD1$^\circ$4381 quite well. 

Synthetic spectra are shown in Figs. \ref{fig:cii}, \ref{fig:dib}, \ref{fig:feii}, and \ref{fig:caii}. Synthetic \ion{Fe}{II} lines from low excitation lines are too deep. Many of these occur in emission in our second spectrum. Most of the \ion{C}{I} lines in synthetic spectra are too strong, which confirms the so-called carbon problem \citep{asplund00}.  
\begin{table}[ht]
\caption{
Chemical composition of the stellar atmosphere.}
\label{tab:chem}
\centering
\begin{tabular}{lr}
\hline\hline
Element  & A    \\
\hline
 H  & <6.00 \\
 He & 12.00 \\
 C  & 10.00 \\
 N  &  8.70 \\
 O  &  9.48 \\
 Ne &  8.48 \\
 Mg &  6.30 \\
 Al &  5.78 \\
 Si &  6.30 \\
 P  &  5.11 \\
 S  &  7.00 \\
 Ca &  5.00 \\
 Ti &  3.85 \\
 Cr &  4.48 \\
 Fe &  6.48 \\
 Ni &  5.48 \\
\hline
\end{tabular}
\end{table}

\section{Interpretation and discussion}
\label{interp}

\subsection{RCB fadings versus eclipses}

From the very beginning, we labeled the two dimming events that occurred in 1949 and 2015 as eclipse-like events. However, we cannot completely rule out the possibility that these are the usual fading events typical of an RCB star. A rule of thumb is that three eclipses are necessary to definitely confirm the eclipses. This is especially true for dusty eclipses, which may change their shapes but not the period. In our case, one will have to wait till 2080 for such confirmation. At present we argue that: typical fading events are 1-9 mag deep with a median depth of about 4.6 mag; they are highly asymmetric with a steep ingress and shallow egress; and only 6\% of isolated fadings are longer than 2 yr \citep{crawford25}. The RCB fading events in the optical are usually accompanied by a response in the IR region, which does not seem to be the case here. Our eclipses would be fairly unusual fading events. 
We would like to note that the Gaia DR3 astrometric fit would not be sensitive to such a long binary star period. Hence, it is no surprise that the star is not included in the Gaia DR3 Part 3. Non-single stars catalog \citep{gaia23}. That is why we proceeded with an analysis as if it were an eclipsing binary star.

To get an idea of the size of the system, we assumed the mass of our RCB star to be about $M_1^\star=1$ M$_\odot$ and that the mass of the companion is similar, i.e., $M_2^\star=1$ M$_\odot$. Given the orbital period of 66 yr, the separation of stars, assuming circular orbits, would be about 20 au. Assuming the minimum masses for the two stars to be about $M_1^\star=0.6$ M$_\odot$ and $M_2^\star=0.1$ M$_\odot$ would shrink their separation to 14 au. Pushing the mass of the companion to the limits and assuming masses of 6 M$_\odot$
and 1 M$_\odot$ for the companion and RCB star, respectively, the separation would be about 31 au.
These values are slightly larger than the size of the dust cloud produced by the RCB star (7 au), which is the second black body causing the IR excess (see Sec. \ref{sed}). 
However, the SED model assumed a spherical and optically thick object. In reality the dust cloud will be optically thin, at least at the edges, and hence larger. It means that the stellar companion will likely be close to the edge of this dust cloud. One might speculate that a spherical dust cloud around the RCB star, which causes the IR excess, overflows the Roche lobe, creating a dusty disk around the companion. 
There is likely an inner hole in the dust cloud due to dust sublimation in the vicinity of such a hot and luminous RCB supergiant. Assuming a sublimation temperature of carbon dust of 1400 K, gray dust opacity, and a luminosity of the RCB star of 4700 $L_{\odot}$, the dust snow line would be located at about 2.7 au.

Assuming the circular orbit of the stars at the separation of $a=20$ au, one would get a mutual velocity of the stars, $v=2\pi a/P$, of about 9 km\,s$^{-1}$. Then the sum of the radii of both components would be $R^{\star}_1+R^{\star}_2=v t_e$ and assuming a duration of the second half of the eclipse of about $t_e=1.7$ yr would result in $R^{\star}_1+R^{\star}_2=3.3$ au. The size of the RCB star is much smaller than that; hence, the radius of the secondary component would be about $R^{\star}_2\approx 3.2$ au. If the occulting object were a star, it would have to be a supergiant and would clearly dominate the SED at longer wavelengths.
For example if its temperature were T=2000\,K {then} its luminosity at such a radius would be 7000 $L_\odot$. We would see it in the SED and in the spectra, but we do not. This means that the occulting object cannot be a star but is most likely a cool dusty disk. Its luminosity seems to be too low to explain the second black body from Sec.\ref{sed}. Given the smooth shape of the egress of the eclipse, the disk is likely optically thin at the edges. Assuming the duration of the egress to be about 4.1 yr (see Sec. \ref{fig:eclipse} or Fig. \ref{fig:eclipse_phase}) would result in a record-breaking disk radius of about 7.9 au or even bigger if the masses of the components were higher than 1 M$_\odot$.
Obviously, the dust disk will be centered around the companion object with the mass $M_2^\star$ and  be supported by its gravity. The dust would likely originate from the RCB star and could be transferred to the Roche lobe of the companion by the stellar wind and radiation pressure.
Assuming the disk radius of 3.2 au means that the relative Hill sphere of the companion $H=R^{\star}_2/a$ is larger than $3.2/20=0.16$. From this, based on the analytical approximation of the effective Roche lobe radius by \cite{eggleton83}, it follows that the mass ratio of the companion must be larger than $M^{\star}_2/M^{\star}_1=0.05$. Given the masses of RCB stars, this means that the companion could also be a massive brown dwarf but a red dwarf star would statistically be the most likely.

The very long orbital period indicates that we are very fortunate to find a system with such an edge-on inclination to observe the eclipses. The probability of eclipses depends on the size of the RCB star, $R^{\star}_1$, and vertical extent of the disk, $H_D$, and is about $(R^{\star}_1+H_D)/a$.
To cause such a deep eclipse, the vertical extent of the disk must be comparable to the size of the RCB star or larger. However, the flat bottom of the eclipse indicates that a small piece of RCB star may still be visible during the eclipse. Hence, assuming that the vertical extent of the disk is about the same as the size of the RCB star and $a=20$ au we get an estimate of the probability of eclipses of about 0.01.
This would mean that there likely are many more non-eclipsing systems of this kind.

\subsection{Model of the dust disk and its transit}
\label{model}
To verify the feasibility of our interpretation, namely that the eclipse is caused by the transit of a huge dusty disk in front of the RCB star, we developed the following model
of the disk and eclipse.
We used the radiative transfer code Shellspec \citep{budaj19} to calculate the light curves. The code was modified to account for dust particles with an optional grain size.
We assumed a dusty disk made of homogeneous spherical carbon grains. Their opacities were the ones of \cite{budaj15}, which are based on the refractive index measurements from \cite{jager98}.
Following on our estimates above, it was assumed that the disk has a radius of 3.2 au and is
orbiting the RCB star on a circular orbit with a separation of about 20 au and an orbital period of about 66 yr.
We assumed the properties of the RCB star from Sect. \ref{sed} ($T_1=11000\,{\rm K}, R_1=19\,R_{\odot}$, $E(B-V)=1.29$ mag).
The limb darkening of the RCB star was treated using the quadratic limb darkening law and the limb darkening coefficients were taken from \cite{claret11}.

The Shellspec code offers a few predefined structures. We used a disk-like structure called the NEBULA. It is composed of gas and dust, but we turned off the gas opacity because it is usually much smaller than the dust opacity. The disk properties are described in the cylindrical coordinates $(r,z)$ aligned with the disk. It has the following radial temperature behavior, which is identical for both gas and dust:  
\begin {equation}
T(r)=T_{in}(r/r_{\rm in})^{etmp} ,
\end {equation}
where $T_{in}$ is the temperature at the inner disk radius, $r_{\rm in}$,
which we put equal to the carbon dust condensation temperature of 1400 K \citep{goeres92}.
Such temperature behavior is expected for irradiated disks with a typical value of $etmp=-0.5$.
The dust temperature and its thermal radiation do not play a crucial role in the optical domain.

Vertical density profiles of gas and dust, as well as the thickness of the disk, are controlled
by the scale height, $H$, which is a function of radius,
Keplerian velocity, $v$, and the speed of sound, $c_{\rm s}$:
\begin {equation}
H(r)=h_{\rm e} \frac{c_{\rm s}(r)}{v(r)}r.
\end {equation}
Such behavior is expected for steady-state circular accretion disks.
We introduced an extra free parameter, $h_{\rm e}$, which might be adjusted to fit the observations, expand the pressure scale height, and for example mimic the effect of turbulence,
chemical composition, or the mass of the central body.
We assumed a circular Keplerian velocity:
\begin {equation}
v(r)=\sqrt{G\frac{M^{\star}_{2}}{r}},
\end {equation}
where $G$ is the gravitational constant and $M^{\star}_{2}$ is the unknown mass in the center of the disk, which we assumed to be $M^{\star}_2=1 M_\odot$.
The speed of sound is
\begin {equation}
c_{\rm s}(r)=\sqrt{\gamma k T(r)/\mu},
\end {equation}
where $\gamma$ is an adiabatic index, $k$ is the Boltzmann constant, and $\mu$ is the mean molecular weight.
We assumed that the surface dust density, $\Sigma_r$, varies as a power law:
\begin {equation}
\Sigma_{r}=\Sigma_{r_{\rm in}}(r/r_{\rm in})^{eden} ,
\end {equation}
where $r_{\rm in},\Sigma_{r_{\rm in}}, eden$ are the inner radius of the disk, a surface density at that point, and a density exponent, respectively.
Further, we assumed the Gaussian vertical density distribution that is
typical of steady-state gas-pressure do\-mi\-na\-ted accretion disks:
\begin {equation}
\rho(r,z)=\rho(r,0)\exp\left(-\frac{z^2}{2H^2}\right) .
\end {equation}
Its mid-plane gas or dust density, $\rho(r,0)$, was determined (as a function of 
the distance) from the surface density and the vertical scale height, 
$H$, using the relation
\begin {equation}
\Sigma_r=\sqrt{2\pi}H(r)\rho(r,0).
\end {equation} 
Our disk has a flared geometry, which means that it gets thicker at larger distances from its center. We assumed that the disk and RCB star suffer the same interstellar extinction as in the Sect. \ref{sed}. During an eclipse of a bright star by dust, it is important to consider the non-isotropic scattering of light because we are in the forward scattering regime (star-dust-observer aligned), which may be very strong \citep{budaj11epsaur}. We took it into account using the dust phase functions from \cite{budaj15}. In this particular case of carbon grains, we found that it is not very important. Once we have such a 3D model of the dusty disk, we let it transit the RCB star. The Shellspec code solves a simple radiative transfer along the line of sight and computes emerging intensities. They are shown in the form of a 2D image in Fig. \ref{fig:eclipse_disk}. These intensities are integrated over the 2D image to get the fluxes. Fluxes as a function of time constitute the light curve.

\begin{figure}[ht]
\centerline{
\includegraphics[width=0.45\textwidth]{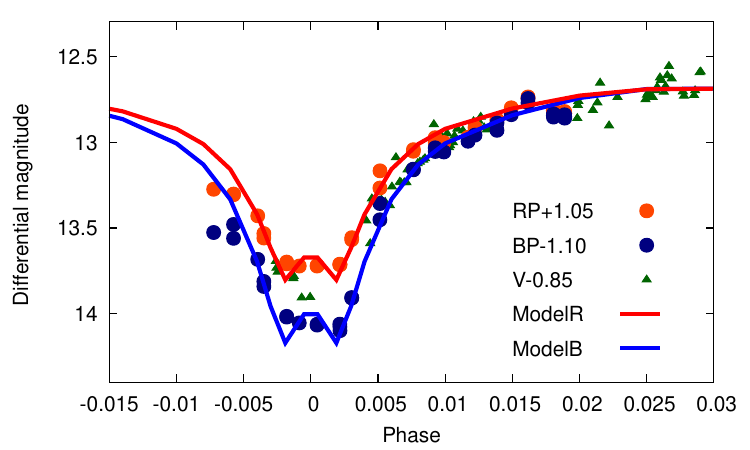}}
\caption{Comparison of the eclipse observations with our model. All observations and models were offset to the same level. Observations are: Gaia/RP-red, Gaia/BP-blue, ASAS-SN/V-green.}
\label{fig:eclipse_model}
\end{figure}

\begin{figure}[ht]
\centerline{
\includegraphics[width=0.49\textwidth]{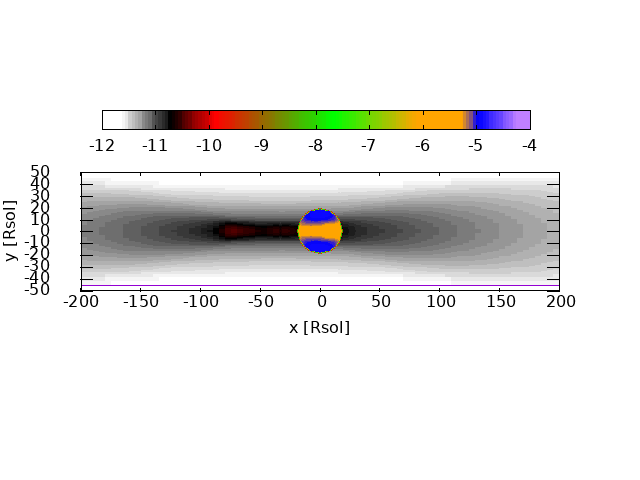}}
\caption{Eclipse of the RCB star (blue circle) at Gaia RP passband and phase -0.0012. The picture zooms on the central part of the disk. Colors are coupled to the logarithm of the observed intensity in erg\,cm$^{-2}$\,s$^{-1}$\,Hz$^{-1}$\,sterad$^{-1}$.}
\label{fig:eclipse_disk}
\end{figure}

We investigated different inner disk radii, mid-plane dust densities, radial density exponents, vertical pressure scale heights, and dust grain radii and their effects on the eclipse light curve.
We found a model that can describe the observed eclipse reasonably well.
The properties of this model are summarized in the Table \ref{tab:disk}, the model light curves are shown in Fig. \ref{fig:eclipse_model}, and the 2D image of the central part of the eclipse is shown in Fig.  \ref{fig:eclipse_disk}.
As can be seen, such disk transits can be used to study the disk and dust properties. For example 
its radial density exponent of 0.95 is in agreement with the values found earlier (1.0-0.6) by \cite{budaj05} and \cite{broz21} for the gas accretion disks around early-type stars. It is slightly smaller than the exponent of the radial surface density profile in the minimum-mass solar nebula model of 1.5 \citep{hayashi81} or the similar exponent derived for other stars with multiple planets based on the so-called minimum-mass extrasolar nebula model (1.6-1.75) \citep{chiang13mmneb,dai20}.
The vertical pressure scale height of the disk depends also on the mass of the body inside the disk and the value of $h_e=2$ indicates that the observed vertical density structure is in reasonable agreement with the steady-state model of the disk. 
Carbon grains with a typical size of about 0.06 micron seem to represent the color dependence of the central part of the eclipse best. The total mass of these grains
in the disk is about $M_D=1\,10^{-12}M_{\odot} (4\,10^{-7} M_{\oplus})$. This is at least 4 orders of magnitude less than the minimum amount of dust causing the IR excess (see Sec. \ref{sed}). Based on the eclipse at optical wavelengths, we cannot exclude the presence of bigger grains that are significantly less active in the optical region. It is apparent that this dusty disk is a different dust structure than the one causing the IR excess, given that its size and mass are insufficient to produce the observed excess. 

One could also propose that the central flat part of the eclipse is a classical total eclipse. Assuming $a=20$ au, this would require a secondary star with a radius of about $80 R_{\odot}$ and brightness in the optical region comparable to the RCB star. We should see such a secondary star in the spectra but we do not. At the same time, it is not in accord with a dimming event that is deeper than the eclipse. We therefore do not consider this total eclipse scenario to be plausible. Nevertheless, the situation may be more complicated with such  a secondary star embedded inside the dark dusty disk. There are so many unknowns about this system, which is why our pure dark dusty disk model should be considered only as a demonstration that works reasonably well.
We certainly do not claim that it is the only possible model. The main unknown is the object in the center of the disk. It could affect the central part of the eclipse significantly.

\begin{table}[ht]
\caption{Parameters of the eclipsing disk and stars.}
\label{tab:disk}
\centering
\begin{tabular}{ll}
\hline\hline
Param.                      &  Value        \\
\hline
\multicolumn{2}{c}{Disk} \\
\hline
$r_{\rm in} [R_{\odot}]$    &  40           \\
$r_{\rm out} [R_{\odot}]$   &  600          \\
$h_{\rm e}$ []              &  2            \\
i [deg]                     &  90.          \\
$T_{in}$ [K]                &  1400         \\
etmp []                     &  -0.5         \\
$\rho (r_{\rm in},0) \rm [g/cm^3]$ & 6.7\,$10^{-18}$   \\
eden []                     & -0.95         \\
$r_{grain}$ [mic]           & 0.06          \\
$M_{D} [M_{\oplus}]$        & $4\,10^{-7}$  \\
\hline
\multicolumn{2}{c}{Stars}                   \\
\hline
$R^{\star}_1 [R_{\odot}]$   & 19            \\
$M^{\star}_1 [M_{\odot}]$   & 1             \\
$T^{\star}_1 [K]$           & 11000          \\
$M^{\star}_{2} [M_{\odot}]$    & 1             \\
a [au]                      & 20            \\
E(B-V) [mag]                & 1.29         \\
\hline
\end{tabular}
\end{table}

\section{Conclusions}

The main conclusions of this work are summarized below.
 
\begin{itemize}
\item 
We searched for eclipse-like events in RCB stars that might be indicative of a binary companion
and found one star of this kind. If confirmed this would be the first binary RCB star.
Unfortunately, the nature of the companion is not known and it could be a star or even a brown dwarf. 
\item     
The star features three major dimming events. An event that occurred in 1999 is an RCB fading event. 
However, a more than 3 yr long dimming event centered on February 2015 is likely an eclipse of the RCB star by a dusty disk associated with the companion. Another dimming event centered on May 1949 is probably the same primary eclipse. This would mean that this RCB star is an eclipsing binary with the second-longest orbital period of about 66 yr and possibly with the largest dust disk. 
\item 
We developed a model of the disk and of the eclipse. The model
can reproduce a ram \aries -like shape of the eclipse, its duration, and its color dependence reasonably well. The dusty disk is made of small carbon grains, its radius is more than 3 au (possibly exceeding 8 au), and contains about $10^{-12}M_{\odot} (4\,10^{-7} M_{\oplus})$ of dust. It was demonstrated that such transiting dusty disks offer us a unique opportunity to determine the radial and vertical structure of the disk.
\item 
The star has a clear IR excess. It is common for RCB stars and is due to their dusty envelopes.
The fact that IR excess is well described by a single black body clearly separated from the radiation in the optical region indicates that IR excess is caused by particles larger than 1 micron while smaller and hotter particles are depleted.  The size of the dusty envelope is larger than 7 au, with the total amount of dust more than $1\,10^{-8} M_{\odot} (3\,10^{-3} M_{\oplus})$. IR brightness is variable with a possible period of about 7 yr.
\item 
The follow-up spectroscopy confirms that the primary source of light is a hydrogen-deficient carbon-rich star. It features strong \ion{He}{I}, \ion{Ne}{I}, and \ion{C}{II} absorption lines. Apart from that, we detected  sharp \ion{Fe}{II} emission lines, which likely originate from a hotter chromosphere-like region. They are variable and accompanied by an underlying broader absorption, indicating that heating of the chromosphere is associated with cooling underneath. We also detected narrow forbidden [OI] emission lines during maximum brightness, which indicate the presence of a vast low-density oxygen-rich nebula. The star features extremely strong DIBs.
\item 
Atmospheric models indicate a strong density inversion in the deeper layers and temperature of about 11000\,K. Its chemical composition resembles some cool EHe stars. 
\item 
We also confirmed exceptionally strong pulsation activity of the star.  
\end{itemize}

With its temperature of 11000\,K, RCB and EHe-like abundances, and the presence of emission lines, this star is bridging the properties and evolution of cool and hot RCBs as well as EHe stars.
It may also offer an important hint about the nature of RCB stars, which are supposed to be merger products, and hence not present in close binary systems.
The probability of such eclipses at such long orbital periods is quite small (about 1\%).
Hence, we encourage researchers to search for other similar non-eclipsing RCB binaries. 
If RCB stars are mergers, then this was originally a triple star. One might speculate and suggest that the existence of a third star is not an accident. It might have absorbed the angular momentum from the inner close binary, shortening its orbital period and thus facilitating formation of the merger. The short period and contact binaries (some of which are precursors of the common envelope phase) are well known to reside in triple or multiple systems and the distant companion might have helped to shorten their orbital periods \citep{pribulla06,tokovinin06}.

\subsection*{Data availability}
The NOT spectra presented in this article are archived at the CDS (unistra.fr) and can be accessed via [].

\begin{acknowledgements}
Authors would like to thank Prof. Geoffrey Clayton for valuable comments that helped to improve the manuscript significantly, Dr. Richard Kom\v{z}\'{i}k for his help with data mining, Dr. John Telting for his advice and info about the FIES instrument, and Prof. Martin Asplund for explaining details of his abundance scale.
JB was supported by the VEGA 2/0033/26 and APVV-24-0160 grants.

This work has made use of data provided by Digital Access to a Sky Century @ Harvard (DASCH), which has been partially supported by NSF grants AST-0407380, AST-0909073, and AST-1313370. Work on DASCH Data Release 7 received support from the Smithsonian American Women’s History Initiative Pool.

Based on observations obtained with the Samuel Oschin Telescope 48-inch and the 60-inch Telescope at the Palomar Observatory as part of the Zwicky Transient Facility project. 
ZTF is supported by the National Science Foundation under Grants No. AST-1440341 and AST-2034437 and a collaboration including current partners Caltech, IPAC, the Weizmann Institute for Science, the Oskar Klein Center at Stockholm University, the University of Maryland, Deutsches Elektronen-Synchrotron and Humboldt University, the TANGO Consortium of Taiwan, the University of Wisconsin at Milwaukee, Trinity College Dublin, Lawrence Livermore National Laboratories, IN2P3, University of Warwick, Ruhr University Bochum, Northwestern University and former partners the University of Washington, Los Alamos National Laboratories, and Lawrence Berkeley National Laboratories. 
Operations are conducted by COO, IPAC, and UW.

This publication makes use of data products from the Wide-field Infrared Survey Explorer, which is a joint project of the University of California, Los Angeles, and the Jet Propulsion Laboratory/California Institute of Technology, funded by the National Aeronautics and Space Administration.
This publication also makes use of data products from NEOWISE, which is a project of the Jet Propulsion Laboratory/California Institute of Technology, funded by the Planetary Science Division of the National Aeronautics and Space Administration.

This work has made use of data from the European Space Agency (ESA) mission
{\it Gaia} (\url{https://www.cosmos.esa.int/gaia}), processed by the {\it Gaia}
Data Processing and Analysis Consortium (DPAC,
\url{https://www.cosmos.esa.int/web/gaia/dpac/consortium}). Funding for the DPAC
has been provided by national institutions, in particular the institutions
participating in the {\it Gaia} Multilateral Agreement.

This publication makes use of VOSA, 
developed under the Spanish Virtual Observatory project supported 
from the Spanish MICINN through grant AyA2008-02156.

This research has made use of the International Variable Star Index (VSX) database, operated at AAVSO, Cambridge, Massachusetts, USA.

IRAF is distributed by the National Optical Astronomy Observatory, which is operated by the Association of Universities for Research in Astronomy (AURA) under cooperative agreement with the National Science Foundation. 

Based on observations made with the Nordic Optical Telescope, owned in collaboration by the University of Turku and Aarhus University, and operated jointly by Aarhus University, the University of Turku and the University of Oslo, representing Denmark, Finland and Norway, the University of Iceland and Stockholm University at the Observatorio del Roque de los Muchachos, La Palma, Spain, of the Instituto de Astrofisica de Canarias. The NOT data were obtained under program ID 71-405.
\end{acknowledgements}

\bibliographystyle{aa} 
\bibliography{budajrcb}

\begin{appendix}

\section{Basic properties of the star}
\label{prop}

\begin{figure}[ht]
\centerline{
\includegraphics[width=0.23\textwidth]{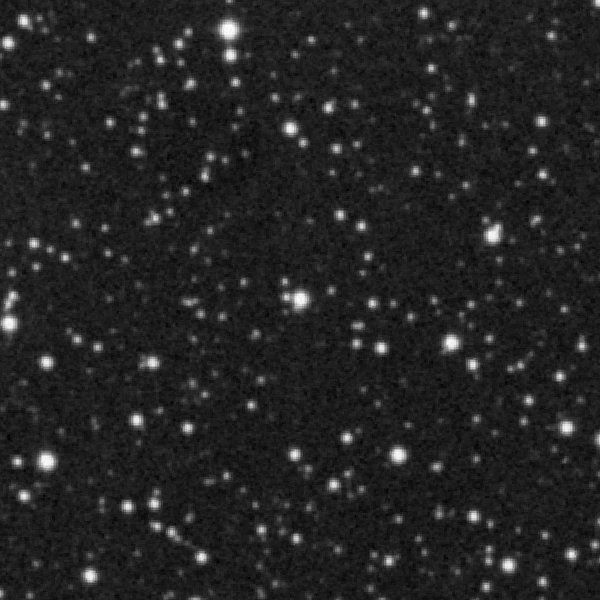}
\includegraphics[width=0.23\textwidth]{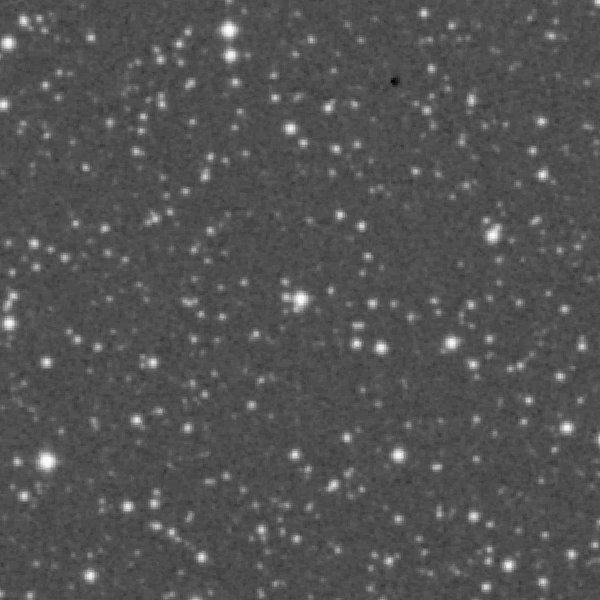}}
\centerline{
\includegraphics[width=0.23\textwidth]{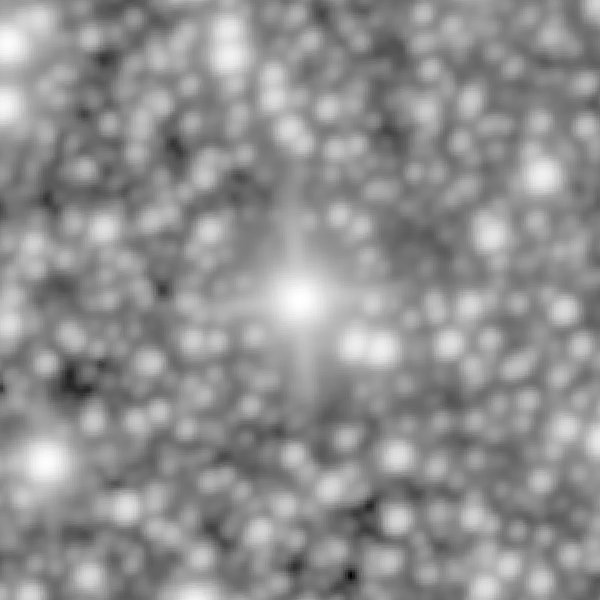}
\includegraphics[width=0.23\textwidth]{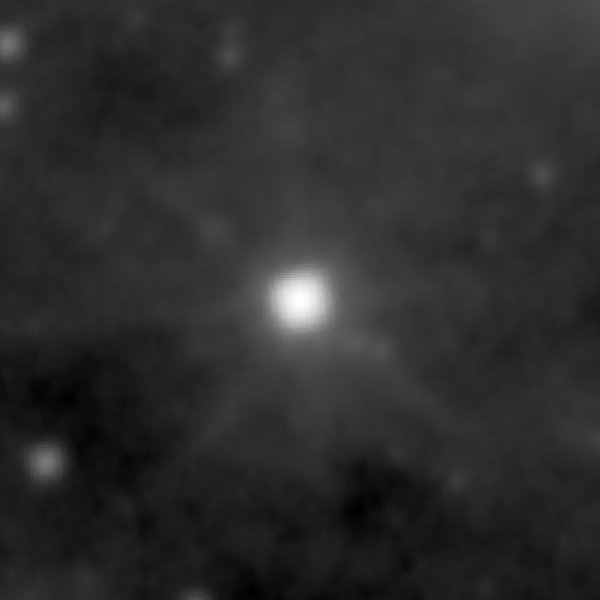}}
\caption{Stellar field in the vicinity of the star in different filters: 
DSS2-blue (top-left), 
DSS2-red (top-right),
WISE-W1 (bottom-left), and 
WISE-W3 (bottom-right). 
Our star is in the center.
Images are 6'x6'. North is up, East is left.}
\label{fig:img}
\end{figure}

\begin{table}[ht]
\caption{
Properties of the star. }
\label{tab:star}
\centering
\begin{tabular}{lrr}
\hline\hline
Property  & Value & Ref.\\
\hline
Gaia ID           & Gaia DR3 524436985982459904  &  1  \\
RA(2016) [deg]    &   12.8669915300   &  1  \\
DEC(2016) [deg]   &  +64.9476987988   &  1  \\
l [deg]           &   122.9351028116  &  1  \\
b [deg]           &   2.0759489845    &  1  \\
pmRA [mas/yr]     & $-1.516\pm0.015$  &  1  \\
pmDEC [mas/yr]    & $ 0.236\pm0.019$  &  1  \\
Gaia G            &  13.145           &  1  \\
Gaia BP           &  14.324           &  1  \\
Gaia RP           &  12.050           &  1  \\
Gaia BP-RP        &   2.274           &  1  \\
$\pi$ [mas]       & $0.155\pm 0.015$  &  1  \\
Dist. [pc]        & $6500 \pm 650$    &  1  \\
Dist. [pc]        & $5300 \pm 500$    &  2  \\
Dist. [pc]        & 5100$\pm$ 1200    &  4  \\
RUWE              & 1.5               &  1  \\
E(g-r)            & 0.96              &  3  \\
A(g)              & 3.377             &  3  \\
A(r)              & 2.512             &  3  \\
E(B-V)            & 0.942             &  3  \\
A(V)              & 2.920             &  3  \\
E(B-V)            & 1.29              &  4  \\
A(V)              & 4.00              &  4  \\
T$_{\rm eff}$ [K]          & 11200 $\pm$ 1000  & 4  \\
$v\sin i$ [km\,s$^{-1}$]   & 15 $\pm$ 5        & 4  \\
\hline
\end{tabular}
\tablefoot{References: (1) -\cite{gaia23}, 
(2) -\cite{bailer-jones21}, 
(3) -\cite{green19},
(4) -this paper.}
\end{table}

\section{More detailed information about the spectra and spectral lines.}

\begin{table}[ht]
\caption{
Information about the spectra. }
\label{tab:spectra}
\centering
\begin{tabular}{lrrr}
\hline\hline
Date        & Time (UT)  & JD &Exposure  \\
yyyy-mm-dd  & hh:mm:ss   & [d]   &[s]       \\
\hline
2025-06-27  & 04:29:10   & 2460853.69 & 3x900     \\
2025-06-29  & 04:20:33   & 2460855.68 & 3x900     \\
 2025-10-29 & 21:55:58   & 2460978.42 & 3x900     \\
\hline
\end{tabular}
\tablefoot{The first two nights are referred to as "1st spectrum" and the 3rd night is referred to as "2nd spectrum".}
\end{table}

\begin{table}[ht]
\caption{
List of selected lines suitable for radial velocity and equivalent width measurements from the 1st spectrum.}
\label{tab:lines1}
\centering
\begin{tabular}{lrrr}
\hline\hline
Wavel.       & EQW  & RV & Ion\\
\hline
   4173.461  &  0.463 & -88.14 & FeII      \\
   4351.769  &  0.464 & -85.49 & FeII      \\
   4387.929  &  0.135 & -89.57 & HeI       \\
   4390.55   &  0.182 & -86.58 & MgII      \\
   4471.498  &  0.324 & -85.95 & HeI       \\
   4481.20   &  0.742 & -88.37 & MgII      \\
   4571.968  &  0.269 & -94.29 & TiII?     \\
   4713.148  &  0.210 & -89.69 & HeI       \\
   4762.433  &  0.307 & -93.35 & CI        \\
   4775.897  &  0.159 & -94.41 & CI        \\
   4876.436  &  0.186 & -93.45 & CrII      \\
   4921.931  &  0.365 & -87.53 & HeI       \\
   5004.195  &  0.098 & -89.08 & FeII      \\
   5015.678  &  0.179 & -88.28 & HeI       \\
   5045.099  &  0.115 & -92.10 & NII       \\
   5052.167  &  0.187 & -95.83 & CI        \\
   5056.14   &  0.440 & -93.62 & SiII      \\
   5216.859  &  0.211 & -89.13 & FeII      \\
   5260.259  &  0.259 & -91.64 & FeII      \\
   5330.735  &  0.330 & -91.84 & OI        \\
   5387.063  &  0.104 & -91.16 & FeII      \\
   5429.988  &  0.164 & -93.64 & FeII      \\
   5436.862  &  0.236 & -92.69 & OI        \\
   5506.195  &  0.214 & -89.84 & FeII      \\
   5554.963  &  0.190 & -94.01 & OI        \\
   5577.31   & -0.759 &  15.43 & [OI]t.e.  \\
   5606.151  &  0.123 & -91.50 & SII       \\
   5875.692  &  0.372 & -89.80 & HeI       \\
   5881.895  &  0.069 & -88.99 & NeI       \\
   5944.834  &  0.120 & -94.91 & NeI       \\
   5978.930  &  0.226 & -90.76 & SiII      \\
   6014.830  &  0.193 & -93.11 & CI        \\
   6016.449  &  0.085 & -93.68 & CI        \\
   6046.388  &  0.254 & -94.35 & OI        \\
   6143.063  &  0.185 & -90.19 & NeI       \\
   6163.594  &  0.104 & -93.19 & NeI       \\
   6300.31   & -0.150 & -92.31 & [OI]s.e.  \\
   6300.31   & -0.090 &  13.04 & [OI]t.e.  \\
   6334.428  &  0.139 & -92.29 & NeI       \\
\hline
\end{tabular}
\tablefoot{The columns are the laboratory wavelength, equivalent widths (both in Angstr\"oms), radial velocity in km\,s$^{-1}$, and chemical element identification (s.e.-stellar emission, t.e.-telluric emission).}
\end{table}

\begin{table}[ht]
\caption{
Continuation of Table \ref{tab:lines1}.}
\label{tab:lines2}
\centering
\begin{tabular}{lrrr}
\hline\hline
Wavel.       & EQW  & RV & Ion\\
\hline
   6347.109  &  0.667 & -86.01 & SiII      \\
   6371.371  &  0.480 & -89.64 & SiII      \\
   6402.246  &  0.264 & -90.89 & NeI       \\
   6484.808  &  0.054 & -91.12 & NI        \\
   6506.528  &  0.187 & -90.31 & NeI       \\
   6532.882  &  0.068 & -89.53 & NeI       \\
   6545.97   &  0.124 & -89.17 & MgII      \\
   6578.052  &  0.583 & -87.14 & CII       \\
   6582.882  &  0.504 & -90.08 & CII       \\
   6598.953  &  0.104 & -91.45 & NeI       \\
   6678.154  &  0.347 & -88.75 & HeI       \\
   6783.907  &  0.110 & -90.77 & CII       \\
   7032.413  &  0.193 & -91.36 & NeI       \\
   7042.083  &  0.306 & -89.78 & AlII      \\
   7056.712  &  0.260 & -96.18 & AlII      \\
   7156.701  &  0.494 & -91.78 & OI        \\
   7231.333  &  0.281 & -88.97 & CII       \\
   7281.349  &  0.104 & -90.17 & HeI       \\
   7423.641  &  0.263 & -93.08 & NI        \\
   7442.298  &  0.345 & -94.50 & NI        \\
   7468.312  &  0.377 & -91.97 & NI        \\
   7771.944  &  0.675 & -85.75 & OI        \\
   7774.166  &  0.569 & -84.49 & OI        \\
   7775.388  &  0.556 & -86.06 & OI        \\
   7877.054  &  0.132 & -91.91 & MgII      \\
   7896.204  &  0.142 & -93.13 & MgII      \\
   7943.164  &  0.273 & -91.11 & OI+NeI    \\
   7947.358  &  0.724 & -89.59 & OI        \\
   8115.225  &  0.090 & -88.25 & MgII      \\
   8133.097  &  0.089 & -89.94 & CaII?     \\
   8184.861  &  0.388 & -95.27 & NI        \\
   8210.715  &  0.303 & -93.84 & NI        \\
   8216.336  &  0.389 & -92.90 & NI        \\
   8242.389  &  0.450 & -90.64 & NI        \\
   8446.     &  0.880 & -      & OI        \\
   8680.282  &  0.649 & -94.60 & NI        \\
   8683.403  &  0.477 & -94.91 & NI        \\
   8686.149  &  0.577 & -84.84 & NI        \\
   8820.423  &  0.603 & -89.73 & OI        \\
\hline
\end{tabular}
\end{table}

\begin{table}[ht]
\caption{
The same as in Table \ref{tab:lines1} but for sharp emission lines measured from the 2nd spectrum.}
\label{tab:emlines}
\centering
\begin{tabular}{lrrr}
\hline\hline
Wavel.       & EQW  & RV & Ion\\
\hline
   5325.55  & -0.055 &  -95.59  & FeII  \\
   5991.38  & -0.142 &  -98.02  & FeII  \\
   6238.39  & -0.102 &  -99.76  & FeII  \\
   6247.56  & -0.118 & -100.24  & FeII  \\
   6416.92  & -0.110 &  -99.04  & FeII  \\
   6432.68  & -0.133 &  -97.78  & FeII  \\
   6516.08  & -0.211 &  -98.69  & FeII  \\
   7449.34  & -0.114 &  -98.03  & FeII  \\
   7462.41  & -0.202 &  -98.10  & FeII  \\
   7515.83  & -0.102 &  -95.53  & FeII  \\
   7533.37  & -0.140 &  -98.33  & FeII  \\
\hline
\end{tabular}
\end{table}

\end{appendix}

\end{document}